\documentclass[journal,9pt]{IEEEtran}

\usepackage{ifpdf}
\ifCLASSINFOpdf
  \usepackage[pdftex]{graphicx}
  \graphicspath{{figures/}}
\else
  \usepackage[dvips]{graphicx}
  \graphicspath{{figures/}}
  \DeclareGraphicsExtensions{.eps}
\fi

\usepackage[caption=false,font=footnotesize]{subfig}

\usepackage{mathtools}
\usepackage{amsmath}
\usepackage{amssymb}
\usepackage{amsxtra}
\usepackage[square, comma, sort&compress]{natbib}
\usepackage{enumerate}
\usepackage{units}
\usepackage{color}
\usepackage{hyperref}
\usepackage{setspace}
\usepackage{multirow}

\DeclareMathOperator*{\argmin}{argmin}
\newcommand\abs[1]{\left|#1\right|}
\newcommand{\nth}[1]{{#1}{\text{th}}}
\newcommand{\re}[0]{\mathrm{Re}}

\newcommand{\CM}[0]{$\mathsf{CM}$}
\newcommand{\CA}[0]{$\mathsf{CA}$}

\begin{document}
%
\title{A Low-Complexity Detection Algorithm for the Primary Synchronization Signal in LTE}
%
%
%
\author{Mohammad~H.~Nassralla, Mohammad~M.~Mansour,~\IEEEmembership{Senior~Member,~IEEE,}
        and~Louay~M.A.~Jalloul,~\IEEEmembership{Senior~Member,~IEEE}
\thanks{Copyright (c) 2015 IEEE. Personal use of this material is permitted. However, permission to use this material for any other purposes must be obtained from the IEEE by sending a request to pubs-permissions@ieee.org.}
\thanks{M.~Nassralla and M. M. Mansour are with the Department of Electrical and Computer Engineering at the American University of Beirut, Lebanon, e-mail: mmansour@ieee.org. L. Jalloul is with Qualcomm Inc., San Jose, CA 95110 USA, e-mail: jalloul@ieee.org}}

\markboth{IEEE~Transactions~on~Vehicular~Technology (to appear)}%
{Nassralla, Mansour, Jalloul: A Low-Complexity Detection Algorithm for the Primary Synchronization Signal in LTE}

\maketitle

%
\begin{abstract}
One of the challenging tasks in LTE baseband receiver design is synchronization, which determines the symbol boundary and transmitted frame start-time, and performs cell identification. Conventional algorithms are based on correlation methods that involve a large number of multiplications and thus lead to high receiver hardware complexity and power consumption. In this paper, a hardware-efficient synchronization algorithm for frame timing based on $K$-means clustering schemes is proposed. The algorithm reduces the complexity of the primary synchronization signal for LTE from 24 complex-multiplications, currently best known in the literature, to just 8. Simulation results demonstrate that the proposed algorithm has negligible performance degradation with reduced complexity relative to conventional techniques.
\end{abstract}\vspace{-0.05in}

%
\IEEEpeerreviewmaketitle

%
\vspace{-0.10in}
\section{Introduction}\vspace{-0.05in}
\IEEEPARstart{T}{he} $3{\text{rd}}$ Generation Partnership Project (3GPP) Long Term Evolution (LTE) standard is based on orthogonal frequency-division multiple-access (OFDMA)~\cite{ref_1}. Upon call initiation, a search procedure performed by the user equipment (UE) is triggered to synchronize its receiver to the transmitting base station (known as eNodeB). During synchronization, the UE receiver acquires the frame starting position, symbol timing, carrier frequency offset, and cell identity information. While frame synchronization aims at detecting the beginning of each frame, it is not restricted to the initial call setup. The UE has to periodically search for neighboring cells for possible handovers. Moreover, due to the susceptibility of OFDMA systems to synchronization errors~\cite{2006_Ai}, the UE should always support a dynamic cell search procedure to update the frame timing and compensate for frequency offset, in order to sustain orthogonality among the subcarriers~\cite{ref_1}.

The primary and secondary synchronization signals (PSS and SSS) are transmitted by the base station for cell identification and frame timing. The physical-layer cell ID is defined as $N_{\text{ID}}^{\text{cell}} \triangleq 3N_{\text{ID}}^{(1)} + N_{\text{ID}}^{(2)}$, where $N_{\text{ID}}^{(1)}{\rm{ = 0,1,}}\cdots{\rm{167}}$ and $N_{\text{ID}}^{(2)} = 0,1,2$ are the three possible sector IDs. The synchronization procedure occurs in two stages. In the first stage, the UE acquires symbol timing, frequency offset and $N_{\text{ID}}^{(2)}$ using the PSS. In the second stage, the UE detects the frame boundary, $N_{\text{ID}}^{(1)}$, and the cyclic prefix (CP) length by using the SSS. Hence, detecting the physical-layer cell ID requires performing a large number of correlations at the UE. Specifically, three matched filters in the time domain are needed to detect the PSS signal~‎‎\cite{2008_Popovic}, and 168 frequency domain correlators are needed to get the SSS signal index~‎‎\cite{2010_Xu}. Correlations can also be done using the fast Fourier transform (FFT) (e.g., see~\cite{2007_Parhi}), but at the expense of an increase in hardware. In~\cite{2010_Shen}, neighboring-cell search for LTE systems based on PSS and SSS were investigated, and sufficient signal metrics for multi-cell
search was derived for various channel conditions.

Extensive work has been done in the literature to reduce the synchronization receiver complexity while attaining good performance. In~\cite{1999_Hsieh}, a low-complexity joint frame synchronization and frequency offset estimation scheme for OFDM systems was investigated. In~\cite{2008_Popovic}, the symmetry property of the PSS signal is exploited to reduce the number of complex multiplications (\CM s) per incoming sample from 65 to 33. The number of correlators is reduced also from 3 to 2 since two of the three PSS sequences are complex conjugates of one another~\cite{2008_Popovic},\cite{2012_Zhang}. A differential correlation scheme that exploits the symmetry in the PSS signal to correlate the first half of the PSS with the second half is presented in~\cite{2010_Dammann}. Additional reduction is achieved in~‎‎\cite{2010_Xu} by making use of the periodicity of the CP to detect the start of each OFDM symbol using lagged autocorrelation followed by PSS search at the beginning of each OFDM symbol. In~\cite{2010_Yang} the complexity is reduced to 24 \CM s using the property that the number of distinct elements in a Zadoff-Chu sequence~\cite{1972_Chu} is no greater than one-third of its length.

\emph{Contributions}: In this work, a hardware-efficient synchronization algorithm based on $K$-means clustering is proposed. The aim is to reduce hardware complexity by reducing the number of distinct PSS samples involved in performing the correlation operation. The $K$-means algorithm clusters the PSS samples into groups, each of which is represented by one \emph{cluster leader}. Consequently, correlating the received samples with the original distinct PSS samples can be approximated with correlations of the received samples with their respective PSS cluster leaders. Hardware savings are achieved by correlating a set of $K'$ received samples with the \emph{same} cluster leader, thus requiring just \unit[1]{\CM} instead of \unit[$K'$]{\CM s}, in addition to $K'-1$ complex additions. Simulations show that we can match the acquisition time performance of the best known method in the literature that uses \unit[24]{\CM s} with our approach using just 8 clusters (i.e. \unit[8]{\CM s}). Tradeoffs between the hardware-complexity and performance are also discussed. We emphasize that the clustering operation is done offline and does not require extra hardware. We also describe the sensitivity of our algorithm to the miss-detection probability with number of clusters. Note that the miss-detection probability in turn is related to the acquisition time.

The rest of the paper is organized as follows. Section~\ref{s:Frame Structure and PSS Design} introduces the frame structure,  PSS properties and their application in synchronization, and reviews existing solutions to reduce hardware complexity. Section~\ref{s:proposed_algo} presents the clustering-based synchronization algorithm together with its complexity analysis. Section~\ref{s:simulations} shows simulation results. Finally, Section~\ref{s:conclusion} concludes the paper.

%
\vspace{-0.065in}
\section{Frame Structure and PSS Design in LTE} \vspace{-0.025in}
\label{s:Frame Structure and PSS Design}
\subsection{Frame structure}\label{s:frame_Structure}\vspace{-0.025in}
In LTE the bandwidth is scalable from \unit[1.4]{MHz} to \unit[20]{MHz} and is allocated in terms of resource blocks (RBs) each spanning \unit[180]{kHz}. The smallest system bandwidth corresponds to 6 RBs‎‎~\cite{ref_1}. The subcarrier spacing is \unit[15]{kHz}. Transmission over the subcarriers is arranged into frames. Within each frame, specific sequences are used for synchronization purposes. A radio frame is \unit[10]{ms} long and is divided into 10 subframes, each consisting of two slots. The slot structure and CP length differ between \emph{normal} and \emph{extended} modes. In the normal CP mode, each slot has 7 OFDM symbols. The first symbol has $N + N_{\text{CP1}}$ samples where $N$ is the number of active subcarriers and $N_{\text{CP1}}$  is the CP length of the first symbol. The other 6 symbols have $N + N_{\text{CP2}}$ samples and $N_{\text{CP2}}$ is their CP length. In extended CP mode, each slot has 6 OFDM symbols. Each symbol has $N + N_{\text{ECP}}$ samples and ${N_{\text{ECP}}}$ is the CP length. The PSS is placed in the last OFDM symbol in subframes 0 and 5 in FDD mode, whereas it is placed in the third OFDM symbol of subframes 1 and 6 in the case of TDD mode~\cite{ref_1}.

The PSS is a Zadoff-Chu (ZC) sequence~\cite{1972_Chu} which belongs to the class of Constant Amplitude Zero Auto Correlation (CAZAC) sequences~‎‎\cite{2008_Guey,1972_Chu}. Such sequences are favored for synchronization purposes because they have constant amplitude and exhibit the useful property that cyclically shifted versions of a ZC are orthogonal to one another. Furthermore, the discrete Fourier transform (DFT) of a ZC sequence is another ZC sequence (see e.g.~\cite{2013_JSPS_Mansour}). A ZC sequence of odd length $L$‎~‎\cite{2008_Guey,1972_Chu} is defined as
\begin{equation}
\label{eq2}
{d_u}\left( n \right) = {e^{ - j\frac{{\pi un(n + 1)}}{L}}},\quad n = 0,1, \cdots, L-1,
\end{equation}
where $u$ is the root index chosen to be relatively prime with respect to the length $L$~\cite{2008_Guey}. It is straightforward to show that an odd-length ZC sequence is symmetric about its center element:
\begin{equation}
{d_u}\left( n \right) = {d_u}\left( {L - 1 - n} \right),\quad n = 0,1,\cdots, L',
\end{equation}
where $L'=(L - 1)/2$, and hence~\eqref{eq2} becomes:
\begin{equation}
\label{eq4}
d_u(n) \!=\! \left\{\!\!\!\!
               \begin{array}{ll}
                 e^{- j\frac{\pi un(n + 1)}{L}}, & \!\!\hbox{$n\!=\!0, 1, \cdots, L'-1$;} \\
                 e^{- j\frac{\pi u(n + 1)(n + 2)}{L}}, & \!\!\hbox{$n\!=\!L',L'\!+\!1,\cdots, L\!-\!2$.}
               \end{array}
             \right.
\end{equation}
In LTE, the DC carrier is not used on the downlink. Hence an odd length ZC is chosen ($L=63$, co-prime to $u$) with the center element punctured and mapped to DC. The index $u$ is chosen to take one of the following three values: $25, 29, 34$. Note that the choice of $u=25$ and $u=L-29=34$ generate ZC sequences which are complex conjugates of one another, i.e.:
\begin{equation}
{d_u}\left( n \right) = d_{L - u}^*\left( n \right).
\end{equation}
Only one of the three different PSS sequences may be sent per sector. Each is distinguished by the index $u$. Thus, each of the three sectors per cell is assigned a specific root index $u$. The desired synchronization signal $s_u(n)$ to be transmitted can then be defined as an OFDM signal using an $N$-point inverse FFT (IFFT) by appropriately mapping the ZC elements in~\eqref{eq4} onto the available subcarriers. Since the subcarrier spacing is \unit[15]{kHz}, $s_u(n)$ can occupy at most $72$ subcarriers, which would typically be generated by an $N\!=\!128$ point IFFT. However, to enable more flexible and reduced complexity detection via matched filtering with lengths shorter than 128 samples, the synchronization signal $s_u(n)$ is defined as OFDM signal with up to $64$ subcarriers, including the DC subcarrier. It can then be detected by a matched filter of length $64$.

The length-$63$ ZC sequence in~\eqref{eq4} is used by puncturing the center DC element and mapping the remaining elements symmetrically around the DC subcarrier as follows. Let $k$ denote the subcarrier index and ${D_u}(k)$ the ZC sequence used at subcarrier $k$. Then,\vspace{-0.05in}
\begin{equation}
\label{eq6}
{D_u}\left( k \right) = \left\{ {\begin{array}{*{20}{l}}
{{d_u}\left( {k + 31} \right),}&{k =  - 31, - 30, \cdots , - 1}\\
{{d_u}\left( {k + 30} \right),}&{k = 1,2, \cdots ,30}
\end{array}} \right.
\end{equation}
Therefore, the discrete time-domain signal is given as\vspace{-0.1in}
\begin{equation}
\label{eq7}\hspace{-0.05in}
{s_u}(n) \!=\! \frac{1}{N}\!\!\sum\limits_{k =  - N/2}^{N/2 - 1}\!\!\!\! {{D_u}\left( k \right){e^{ - j2\pi nk/N}}},~ n = 0,1,\cdots, N - 1,
\end{equation}\\[-1.0em]
where $N$ is the size of IFFT block. A CP is added to~\eqref{eq7} before transmission. Note that the PSS is generated using equation~\eqref{eq7}.

%
\vspace{-0.05in}
\subsection{Channel Effect on Transmitted PSS}
The signal $s_u(n)$ is transmitted over a multipath fading channel and corrupted by additive white Gaussian noise (AWGN). The received time-domain baseband signal $r(n)$ after removing the CP in an OFDM symbol is given by\vspace{-0.05in}
\begin{equation}
\label{eq8}
r(n) = {e^{\frac{{^{j2\pi n\varepsilon }}}{N}}}\sum\limits_{m = 0}^{M - 1} {h(m)s_u(n - \theta  - m)}  + z(n),
\end{equation}\\[-1.0em]
for $0 \le n \le L - 1$, where $L$ is the number of observation samples, $h(m)$ is the $\nth{m}$ complex coefficient of the discrete channel impulse response with $M$ resolvable paths, and $z(n)$ is contribution of the thermal noise which is modeled as a zero-mean complex Gaussian circularly symmetric with variance ${\sigma ^2}$. The parameter $\theta$ represents the timing offset. Note that $r(n)$ contains a frequency offset relative to the transmitted signal $s_u(n)$. The normalized carrier frequency offset (CFO) with respect to the local reference is given by $\varepsilon = N{T_s}{f_{\text{CFO}}}$, where $N$ is the IFFT size, $T_s$ is the sampling period and ${f_{\text{CFO}}}$ is the CFO in Hz.


%
\vspace{-1mm}
\subsection{Matched-Filter Detection}
PSS detection is accomplished by maximizing the cross-correlation between ${s_u}(n)$ and $r(n)$. To perform the correlation operation, $r(n)$ is match-filtered with respect to ${s_u}\left( n \right)$ followed by taking magnitude-squared as follows:
\begin{equation}
\label{eq9}
{y_u}\left( m \right) = \left| {\sum\limits_{n = 0}^{N - 1} {r\left( {n + m} \right)s_u^*\left( n \right)} } \right|^2,\;m = 0,1, \cdots N - 1,
\end{equation}
where  $u = 25, 29, 34$.

Since the cross-correlation property of PSS is minimal for all shifts and is maximal for non-zero shifts, then the beginning of PSS can be determined by setting a threshold $\lambda$ to compare $y_u(m)$ against. When the value of $y_u(m)$ exceeds $\lambda$, we then consider that a PSS has been detected and its position is indicated by $m$.
A brute force implementation of~\eqref{eq9} requires $N+1$ complex multiplies and $N\!-\!1$ complex add (\CA) operations per correlation operation (summation and then magnitude). For all values of $m$ and $u$ , the total number of operations hence becomes \unit[$3N(N\!+\!1)$]{\CM} and \unit[$3N(N\!-\!1)$]{\CA} operations, which is excessively high. Two symmetry properties between the elements of the underlying ZC sequences, as well as between the sequences corresponding to distinct root indices, can be exploited to reduce the number of \CM s~\cite{2008_Popovic}. Specifically, since~\eqref{eq6} satisfies
\begin{equation}
\label{11}
{D_u}\left( k \right) = {D_u}\left( {N - k} \right),\quad k = 1, \cdots, N - 1,
\end{equation}
then it follows that ${s_u}\left( n \right)$ in~\eqref{eq7} is also centrally symmetric:
\begin{equation}
{s_u}\left( n \right) = {s_u}\left( {N - n} \right),\quad n = 1, \cdots ,N - 1.
\end{equation}
Second, since the root indices $u$ and $v=L-u$ generate complex-conjugate ZC sequence pairs, then it follows that the corresponding synchronization signals ${s_u}\left( n \right)$ and ${s_v}\left( n \right)$  are complex conjugates as well:
\begin{equation}
{s_u}\left( n \right) = s_{L - u}^*\left( n \right).
\end{equation}
\noindent Therefore in~\eqref{eq9}, the same \CM~operations can be used to compute both ${y_u}\left( m \right)$ and ${y_{L - u}}\left( m \right)$. If $r(n+m) s_u(n) = (a+jb)(c+jd) = (ac-bd)+j(bc+ad)$ for some $a,b,c,d\in\mathbb{R}$, then $r(n+m)s_u^*(n)=(a+jb)(c-jd) = (ac+bd)+j(bc-ad)$. Combining the above simplifications, the total number of operations required to compute ${y _u}\left( m \right)$ for all three indices reduces to $N\cdot\left(2\left(1 + \frac{{N - 1}}{2} \right) + 1\right) = N(N+2)$ \CM~operations and $3N\cdot\left( 1 + \frac{N - 1}{2} + \frac{N - 1}{2} - 1 \right) = 3N(N-1)$ \CA~operations.

%
\section{Proposed Clustering-Based Algorithm}\label{s:proposed_algo}
The main idea in this paper is to partition the PSS samples $s_u(n)$ into clusters such that samples that are similar in the least-squares sense are grouped into the same cluster~\cite{2010_Na,1982_Lloyd}. Each cluster would then be characterized by a cluster leader that represents all the PSS samples in the cluster. These cluster leaders can then be used to perform the correlation operation with the received samples $r(n)$ (instead of using the actual PSS samples as was done in~\eqref{eq9}). Hence, the complex multiply-and-add operation in~\eqref{eq9} between the various PSS samples $s_u(n)$ in the same cluster and their corresponding received samples $r(n)$ at sample-time $n$ reduces to a \emph{single complex multiplication between the cluster leader and the sum of the corresponding received samples} $r(n)$. This results in a significant reduction in complex multiplications. A trade-off exists between these savings and detection accuracy depending on the clustering algorithm employed and desired number of clusters. In the limiting case where one cluster represents only one PSS sample, the correlation operation is identical to~\eqref{eq9}. We emphasize that clustering is an off-line operation that is done only once and does not require any hardware resources. This is unlike the correlation step which has to be repeated in every PSS detection phase. So the correlation step is considered the performance bottleneck and not the clustering step.

%
\vspace{-0.125in}
\subsection{Clustering of PSS Signal into $K$ Clusters}\vspace{-0.05in}
There are several clustering algorithms that can be employed to cluster the PSS samples (e.g., see~\cite{2010_Rokach} for a survey). We use the method of $K$-means clustering to partition the $N$ PSS samples into $K$ clusters such that each PSS sample belongs to the cluster with the nearest mean, serving as a cluster leader. This results in a partitioning of the sample space into Voronoi cells~\cite{1967_MacQueen}. Lloyd's $K$-means algorithm~\cite{1982_Lloyd} partitions the $N$ samples $\mathcal{P}=\{s_u(n)~|~n=0,1,\cdots,N-1\}$ into $K$ sets $\mathcal{P}_1,\cdots,\mathcal{P}_K$ with $K \leq N$, so as to minimize the weighted within-cluster sum-of-squares (WWCSS) as follows:
\begin{align}
\label{eq:wcss_sum}
   \argmin_{\mathcal{P}_1,\cdots,\mathcal{P}_K}~ \sum_{k=1}^{K}w_k\!\!\!\!\sum_{s_u(n)\in \mathcal{P}_k}\!\!\!\| s_u(n)-\mu_k\|^2,
\end{align}
where $\mu_k$ is the mean of the samples in $\mathcal{P}_k$, and $w_k\!\in\! \mathbb{R}$ is a weighting factor. Euclidean distance is used as a clustering metric, and variance is used as a measure of cluster scatter. The final clusters $\mathcal{P}_k$ are non-empty, disjoint, and form a partition, i.e. $\mathcal{P}\!=\!\cup_{k=1}^{K}\mathcal{P}_k$. Let $N_k$ be the number of samples in $\mathcal{P}_k$, i.e., $N_k = \abs{\mathcal{P}_k}$.

It is well known that finding the optimal solution for~\eqref{eq:wcss_sum} is NP-hard~\cite{2009_Aloise}. The Lloyd algorithm uses an iterative refinement technique to approximate~\eqref{eq:wcss_sum} as described next. Starting from an initial arbitrary set of $K$ means, say $\mu_1^{(1)},\cdots,\mu_K^{(1)}$, the algorithm proceeds by alternating between two steps: (1) At step $t$, assign each sample to the cluster whose mean yields the least WWCSS:
\begin{align}
\label{eq:lloyd_assignment_Step}\hspace{-0.1in}
\mathcal{P}_k^{(t)} \!=\! \{s_u(n):\! w_k\| s_u(n)\!-\!\mu_k^{(t)}\|^2 \leq w_j\| s_u(n)\!-\!\mu_j^{(t)}\|^2, \forall j\in[1,k]\},
\end{align}
where each $s_u(n)$ is assigned to exactly one $\mathcal{P}_k^{(t)}$, even if it can be assigned to more than one; (2) The means are updated to be the centroids of the samples in the new clusters:
\begin{align}
\label{eq:lloyd_update_Step}
\mu_k^{(t+1)} = \frac{1}{|\mathcal{P}_k^{(t)}|} \sum_{s_u(n)\in \mathcal{P}_k^{(t)}}\!\!\!s_u(n),\quad k=1,\cdots,K.
\end{align}
The algorithm converges when the assignments no longer change. Since the arithmetic mean is a least-squares estimator, it also minimizes the WWCSS objective. Also since there only exists a finite number of partitions, the algorithm must converge to a (local) optimum, though not necessarily the global optimum.

\begin{figure}[hbtp]
\centering
\includegraphics[width=2.8in]{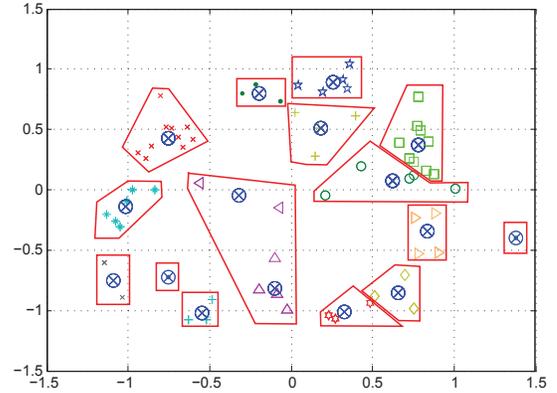}\vspace{-0.05in}
\caption{Clustering of PSS samples into $K=16$ clusters.}
\label{fig:clustering_of_PSS}
\end{figure}

Figure~\ref{fig:clustering_of_PSS} shows the result of clustering the PSS samples into $K=16$ clusters with uniform weighting factors $w_k\!=\!1$, where cluster leaders are marked as {\color{blue}{$\mathbf{\otimes}$}}. The real part of the PSS of~\eqref{eq4} and the clustered PSS can be plotted as well. It can be easily shown that the clustered signal tracks the original signal with high accuracy. A similar behavior applies to the imaginary part. The relevant plots are omitted due to space limitations. Detection performance of clustered PSS into 16 clusters produces a result that is close to the optimum outcome as we will demonstrated through simulation.

Note that the Lloyd algorithm needs to be applied only for PSS signals with $u\!=\!25,29$, since signals with $u=29, 34$ are complex-conjugate of one another, and hence their clustered versions are also complex-conjugates. Note that clustering is done off-line, and the $128\times 2$ cluster indices of the $N=128$ PSS samples, denoted by $\pi_{u}[n]$ for both $u=25$ and $u=29$ and $n=0,1,\cdots,N-1$, are stored in a look-up table (LUT) that is available at the receiver. For example, if a PSS sample maps to sample $n_k$ in cluster $\mathcal{P}_k$ for $u=25$, then the time sequence index of the received sample is retrieved as $\pi_{25}\left(n_k+\sum_{j=1}^{k-1}N_j\right)$.
\vspace{-0.125in}

%
\subsection{Optimized Correlation Using Clustering}
When performing correlation, the samples that correspond to the same cluster can first be added before multiplication. Hence, instead of using $N$ complex multiplications as in~\eqref{eq9}, the $N_k$ samples of $r(n)$ whose corresponding PSS samples belong to the same cluster can be added prior to multiplication. So~\eqref{eq9} can be approximated as
\begin{equation}
\label{eq19}\hspace{-0.05in}
y_u(m)\!\approx\! \abs{\sum_{k = 1}^{K} \mu_{k}^* \sum_{n = 0}^{N_k-1} r\!\left(\!\pi_u\!\!\left[n+\sum_{j=1}^{k-1}N_j\right] \!+\! m \!\!\!\!\pmod{N}\!\right)}^2\!,\!
\end{equation}
where $\mu_{k}^*$ is the complex conjugate of the mean of the $\nth{k}$ cluster, and $\pi_u[\cdot]$ is a LUT that stores the mapping of the cluster indices of the PSS samples. The inner summation accumulates all the received samples belonging to the $\nth{k}$ cluster, and the sum is multiplied by the complex-conjugate of the cluster-mean $\mu_k^*$.

\begin{figure}[hbtp]
\centering
\includegraphics[width=2.8in]{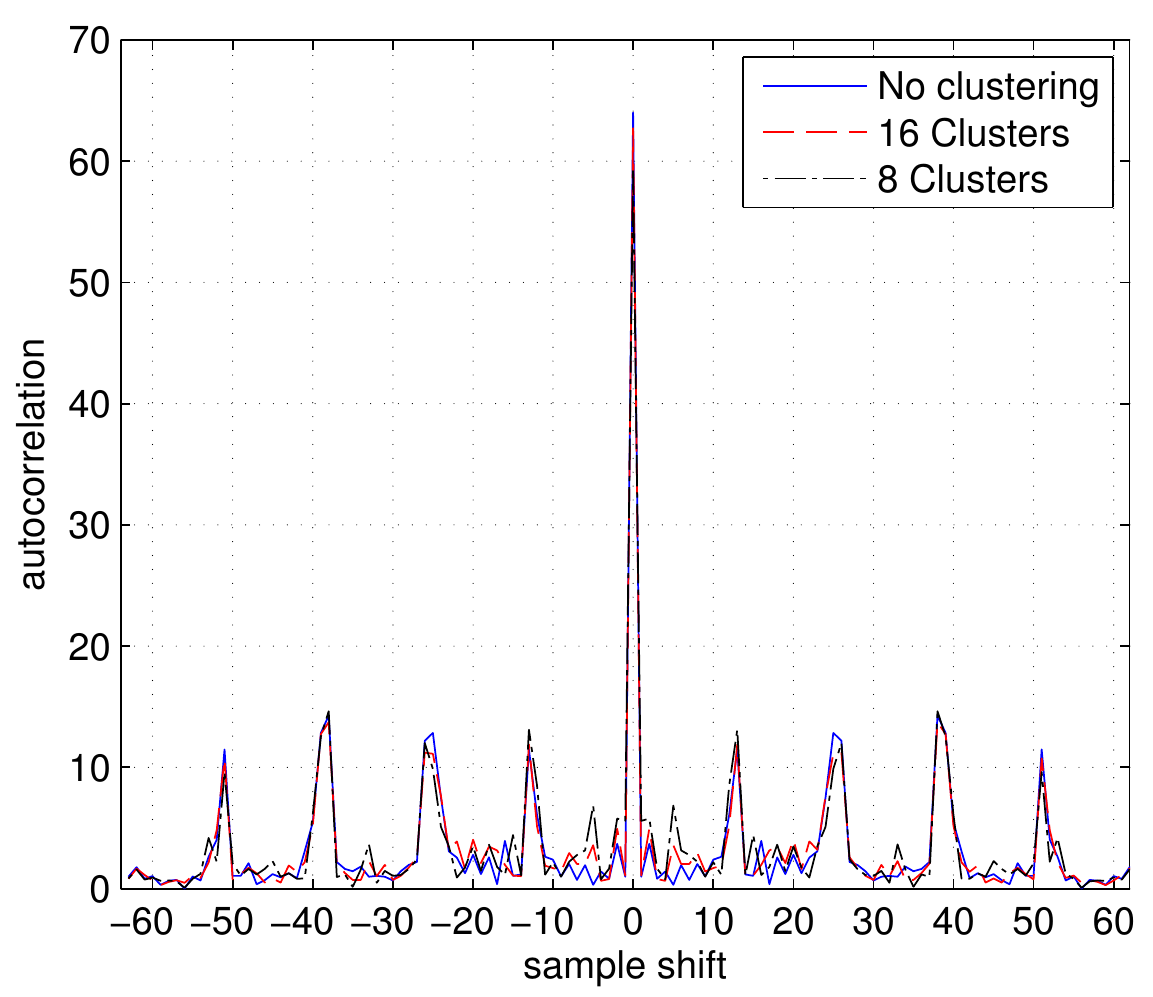}\vspace{-0.125in}
\caption{PSS autocorrelation with and without clustering.}
\label{PSS_autocorr_ZC}
\end{figure}

Figure~\ref{PSS_autocorr_ZC} plots the autocorrelation of the PSS signal without clustering and with clustering into $K=8$ and 16 clusters. The plot clearly shows that the zero autocorrelation property of the ZC sequence is still preserved under clustering.

Note that when the number of clusters equals the length of the PSS sequence (i.e., there is a single element in each cluster), the correlation output in~\eqref{eq19} becomes equivalent to that of the matched filter. This condition on clustering sets an upper bound on the performance of the proposed clustering scheme, but obviously losses the complexity advantage over the matched filter. For any other case wherein the cluster has more than one element, then the correlation term in~\eqref{eq19} will include the matched filter output plus an additional ``interference" term $I_u(m)$ that increases the miss-detection probability:
\begin{align*}
I_u(m) \!&=\! \abs{\sum\limits_{n = 0}^{N - 1} {r(n+m)\left[\mu_{k(n)}^*\!-\!s_u^*(n)\right]} }^2\\ \!&+\!2\re{\left[\left(\sum\limits_{n=0}^{N-1}{r(n+m)s_u^*(n)}\right)\!\!
                 \left(\sum\limits_{n=0}^{N-1}{r(n+m)\mu_{k(n)}^*}\right)^{\!\!*} \right]},
\end{align*}
for $m \!=\! 0,1, \cdots N \!-\! 1$, where $k(n)$ is the index of the cluster to which $s_u(n)$ belongs.

An exact analysis of the miss-detection probability requires the distribution of this additional interference term $I_u(m)$, which is mathematically cumbersome to determine. We therefore resort to computer simulations to determine the miss-detection probability as discussed in Section~\ref{s:simulations}.

%
\subsection{Cluster-Based Correlator Architecture}
\begin{figure}[hbtp]
\centering
\subfloat[Without cyclic shifting]{\includegraphics[width=3in]{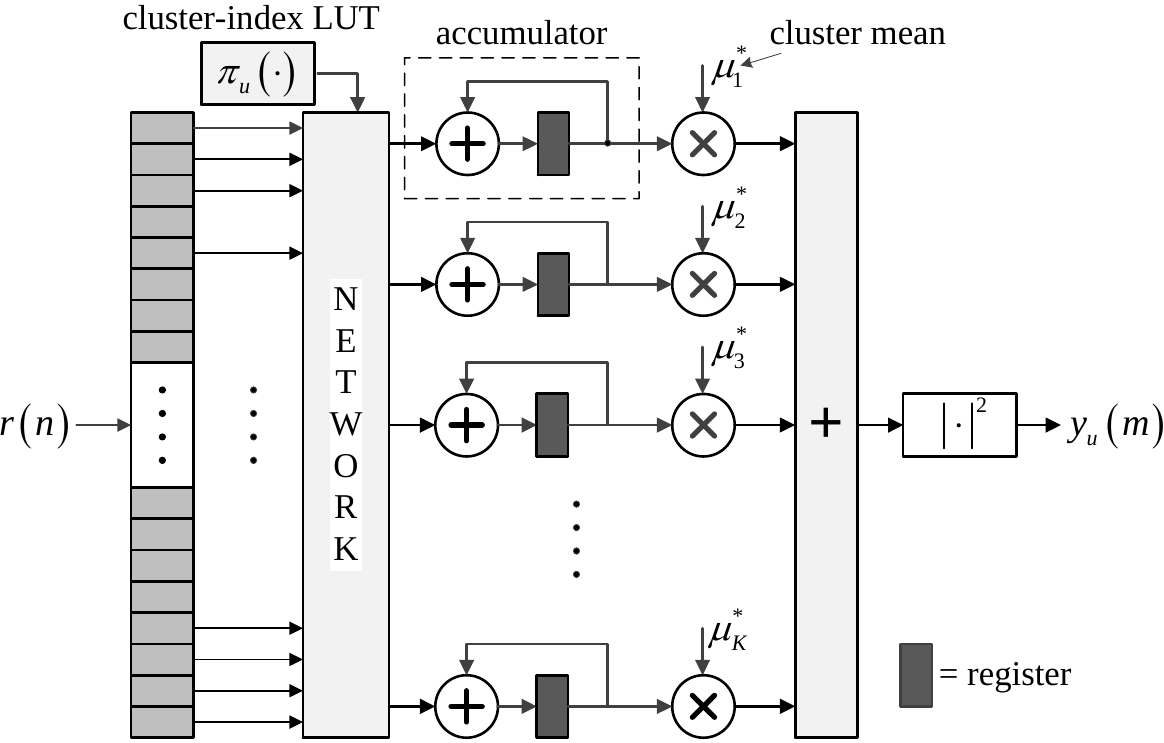}
\label{fig:lloyd_correlator_noshifting}}\\\vspace{-0.1in}
\subfloat[With cyclic shifting]{\includegraphics[width=2.8in]{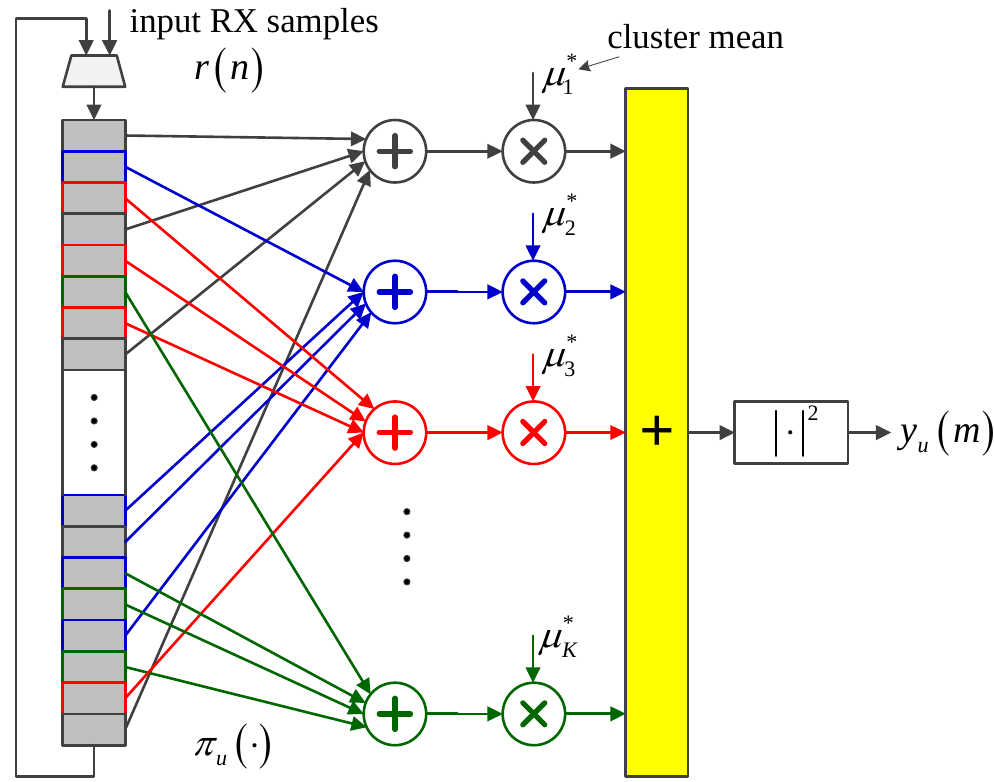}
\label{fig:lloyd_correlator_shifting}}
\caption{Architectures of cluster-based correlators: (a) Using a network with configurable connections to route the $r(n)$ values to their respective clusters for accumulation without actually cyclically shifting the data with every $m$, and (b) Using a shift-register that shifts the data but with hard-wired connections to the adders.}
\label{fig:lloyd_correlator}
\end{figure}

Figure~\ref{fig:lloyd_correlator_noshifting} shows a block diagram of a cluster-based correlator. It employs $K$ complex accumulators and $K$ complex multipliers corresponding to the $K$ clusters. The received samples are steered using the network and $\pi$-LUT to the accumulators of the corresponding clusters. After accumulating all samples in a given cluster, the sum gets multiplied by $\mu_k^*$. The results from all cluster multipliers are then added, and their squared-norm is computed. The steps are repeated for every shift value $m$ without actually shifting the data $r(n)$ in order to minimize data movement and hence reduce switching power. Proper steering of data is handled in this case using the $\pi$-LUT which stores the mapping indices. Alternatively, Fig.~\ref{fig:lloyd_correlator_shifting} shows an architecture that employs a shift-register to move the data for every $m$. The network simplifies to hard-wired connections to the adders before multiplying by $\mu_k^*$, but cyclic data movement of $N$ elements for every $m$ results in increased power consumption. Using the final outputs $y_u(m)$ for $m=0,1,\cdots,N-1$, PSS detection is performed by selecting the maximum autocorrelation value $\mathsf{max}\{y_u(m),m=0,1,\cdots,N-1\}$, and comparing it to a predetermined detection threshold $\lambda$.

A tradeoff exists between complexity and detection accuracy depending on the number of clusters employed. However, if the original number of received samples is small, detection using a small number of clusters can result in a degraded performance. For example, if the received P-SCH signal has only $64$ samples, clustering of the PSS signal into $8$ clusters (to reduce \CM s) and performing correlation accordingly shows a poor performance. A simple fix to this problem is by up-sampling the received signal $r(n)$ by a factor of 2 to generate $N=128$ samples $r'(n)$, where $r'(2n)=r(n)$ and $r'(2n+1)=(r(2n)+r(2n+2))/2$. Clustering of the PSS signal into $8$ clusters is performed always assuming it has 128 samples long as before. It is interesting to note that simulation results show that correlation based on $r'(n)$ only degrade mildly if the number of clusters is reduced from 8 to only 6.

%
\vspace{-0.075in}
\section{Performance Analysis and Simulation Results}\label{s:simulations}
The PSS is received in \unit[1.4]{MHz} bandwidth based on a \unit[1.92]{MHz} OFDM sampling rate. The brute-force optimal matched-filter (MF) detection requires \unit[64]{\CM s} for each incoming sample for each root index $u$. Popovic~\cite{2008_Popovic} and Yang~\cite{2010_Yang} exploited the symmetry in the PSS which resulted in a reduction of MF complexity to 33 and \unit[24]{\CM s}, respectively. Using the proposed algorithm in this paper, only 8 or \unit[6]{\CM s} are required to reach detection performance equivalent to \cite{2008_Popovic}.

The key performance indicator for synchronization, as used in \cite{2008_Popovic}, is the acquisition time which is defined as the time (in msec) it takes for the receiver to acquire symbol timing and cell identification. However, in order to gain more insight into the proposed synchronization algorithm, we first demonstrate its performance through the evaluation of the miss-detection probability and compare it to the MF detection method in \cite{2008_Popovic}. The targeted probability of false alarm is 0.1. The system simulation parameters are as follows: channel model static AWGN or TU 6-path; $\text{UE speed}=\unit[3]{km/h}$; $\text{Carrier frequency} = \unit[2]{GHz}$; 1 receive antenna at the UE; $\text{SNR} = \unit[-5]{dB}$ for user at the cell edge; frequency offset of $\unit[5]{ppm}$ for initial cell search, and $\unit[0.1]{ppm}$ for neighbor cell search; probability of false alarm of 0.1.

\begin{figure}[t]
\centering
\includegraphics[width=3.5in]{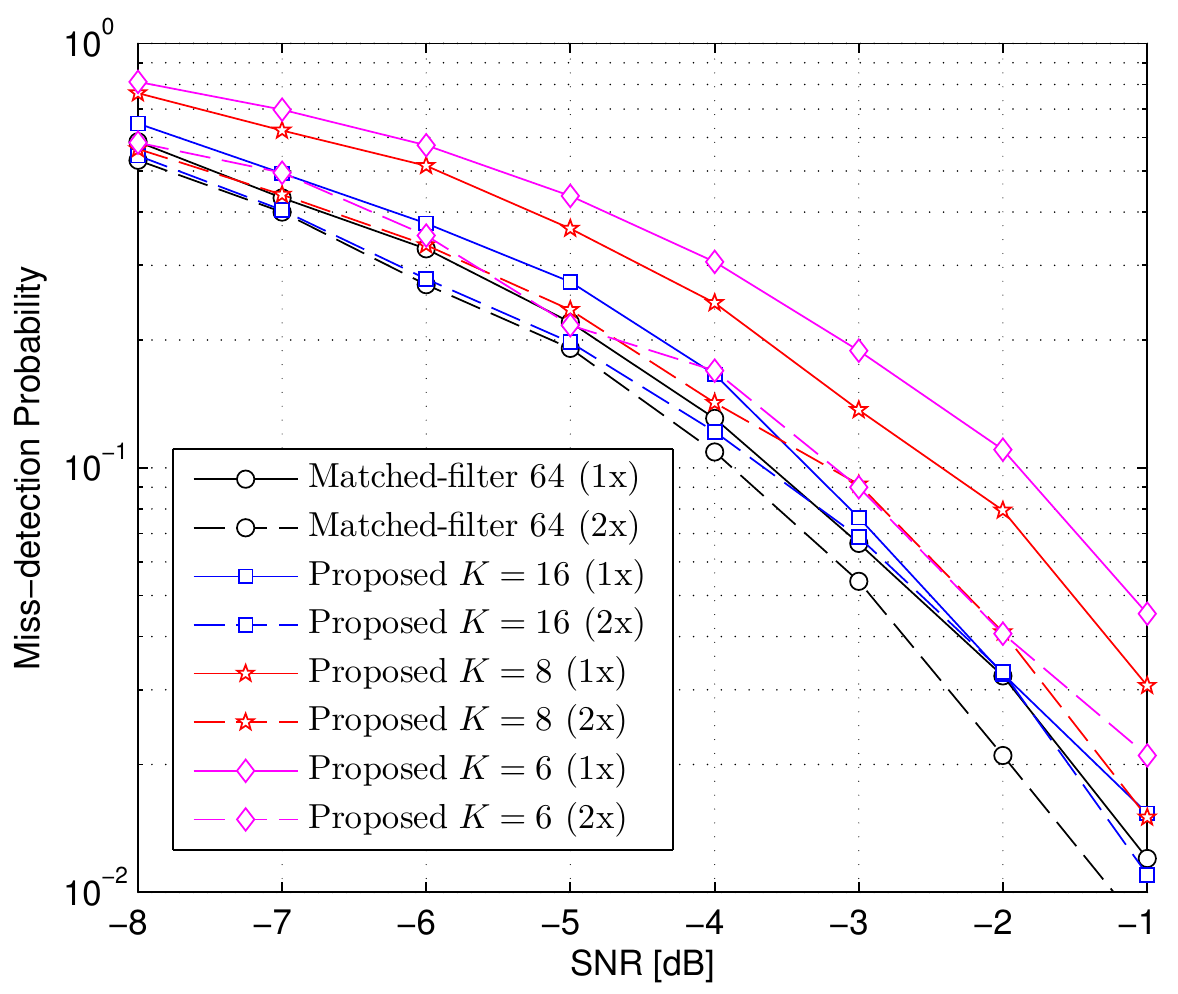}\vspace{-0.125in}
\caption{Probability of miss-detection versus $\text{SNR}$ in dB for a false alarm probability of 0.1.}
\label{Fig:Pmd}
\end{figure}

\begin{figure}[hbtp]
\centering
\includegraphics[width=3.5in]{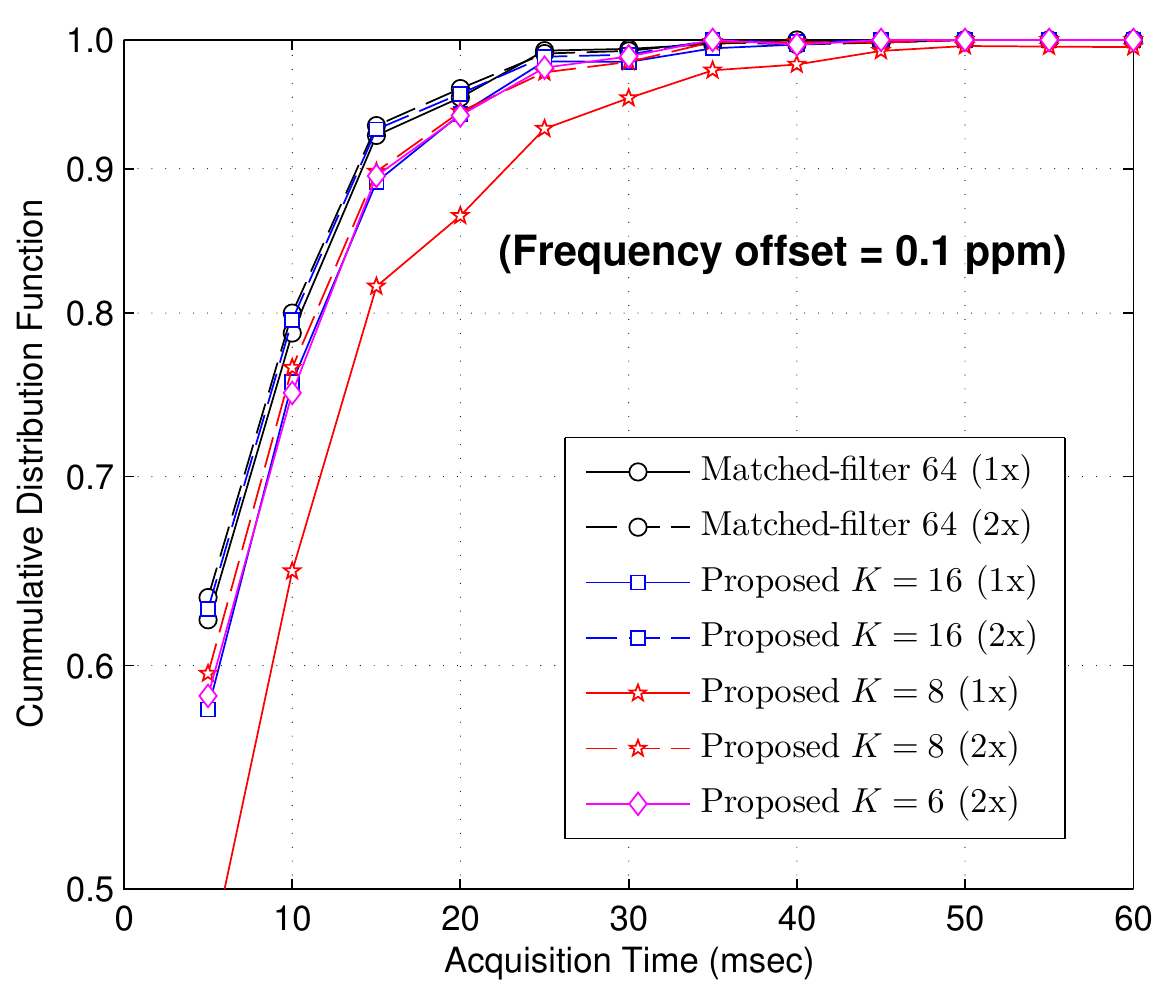}\vspace{-0.125in}
\caption{CDF of acquisition time for $\text{SNR}\!=\!\unit[-5]{dB}$ at \unit[0.1]{ppm} and false alarm
probability of 0.1.}
\label{f:acquisition_time_0pt1ppm}
\end{figure}

\begin{figure}[hbtp]
\centering
\includegraphics[width=3.5in]{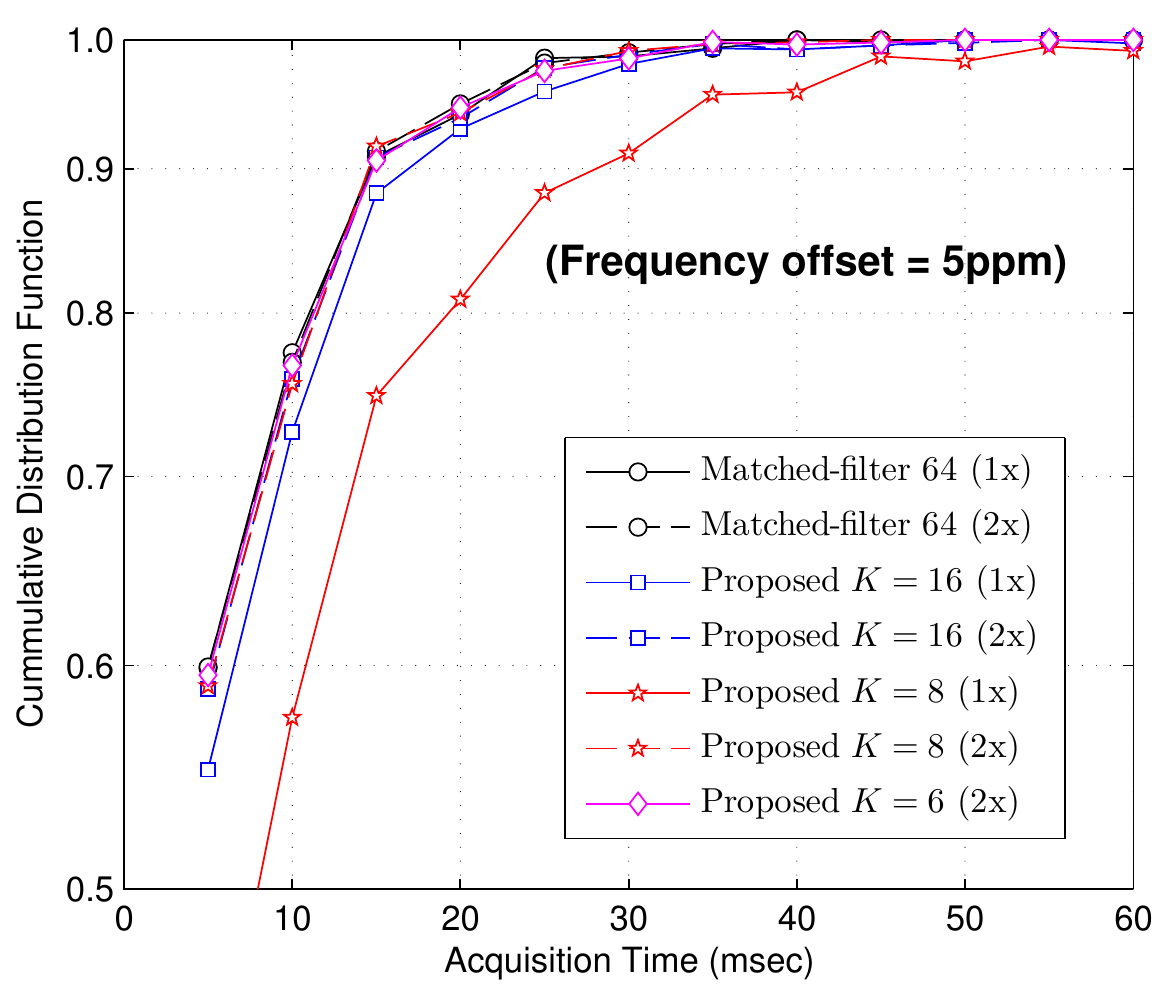}\vspace{-0.125in}
\caption{CDF of acquisition time for $\text{SNR}\!=\!\unit[-5]{dB}$ at \unit[5]{ppm} and false alarm
probability of 0.1.}
\label{f:acquisition_time_5ppm}
\end{figure}

Fig.~\ref{Fig:Pmd} shows the miss-detection probability as a function of SNR for a constant false alarm probability in a static single-path AWGN channel without frequency offset. The proposed method is compared to the MF method of \cite{2008_Popovic} for various cluster sizes and oversampling factors. As seen from Fig.~\ref{Fig:Pmd}, the case of 8 clusters with an oversampling factor of 2 has a \unit[2.5]{dB} SNR improvement over the MF result using 64 samples and is only \unit[0.2]{dB} worse than the MF with 2 times oversampling. Therefore, it is evident that the proposed clustering method shows excellent performance versus complexity tradeoff. This property is manifested in the acquisition time performance which is a useful measure of the amount of cycles it takes to detect the correct PSS sequence. For each instance of the channel and noise realizations, the number of cycles to detect the correct PSS sequence is a random variable. In the simulation results of Figs.~\ref{f:acquisition_time_0pt1ppm} and \ref{f:acquisition_time_5ppm}, the CDF of the acquisition time is plotted using the TU 6-path model for a frequency offset of 0.1 and $\unit[5]{ppm}$, respectively.

The following synchronization detection methods are compared: matched-filter using $1\times$ and $2\times$ oversampling, and the proposed algorithm with $K=16,8,6$ using $1\times$ and $2\times$ oversampling.

Figure~\ref{f:acquisition_time_0pt1ppm} shows a comparison of the CDF of the acquisition times with $\unit[0.1]{ppm}$ of the MF detection as compared to the proposed algorithm which is parameterized by the oversampling factor and the cluster size. It can be seen that the proposed algorithm with $K=16$ and $2\times$ oversampling yields performance that closely aligns with the MF and for $K=16$ and $2\times$ oversampling yields performance within \unit[0.5]{msec} of acquisition time as the MF. However, the proposed algorithm has the advantage of a reduction in the number of \CM s over the MF implementation. Figure~\ref{f:acquisition_time_5ppm} shows a similar trend in the acquisition performance of the proposed method as compared to the MF with a higher frequency offset of $\unit[5]{ppm}$.

As for complexity, the MF implementation requires 65 (2x) and 33 (1x) complex multiplications. However, the proposed K-means clustering algorithm requires 16 complex multiplications for both 2x and 1x oversampling using 16 clusters, and 8 complex multiplications using 8 clusters. Using 6 clusters only, the proposed scheme requires 6 complex multiplications.

%
\vspace{-0.125in}
\section{Conclusions}\label{s:conclusion}\vspace{-0.05in}
A novel clustering-based algorithm for frame synchronization in LTE using the PSS signal has been presented. Clustering the PSS signal into smaller sets represented by a cluster leader using Lloyd's $K$-means algorithm has been shown both to reduce the computational complexity of the synchronization via sample correlation, as well as improve detection performance compared to state-of-the-art. Moreover, up-sampling the received signal to improve performance without increasing the complexity when the PSS has less than 128 samples has been shown to be effective for large frequency offset scenarios. The impact of adapting the weights of the clusters when performing $K$-means clustering will be investigated in a future work.
\ifCLASSOPTIONcaptionsoff
  \newpage
\fi

\vspace{-0.075in}
\footnotesize
\singlespacing
\bibliographystyle{IEEEtran}
\bibliography{IEEEabrv,PSS_Bibliography}

\end{document}